\documentclass[aps,pra,twocolumn,showpacs,nofootinbib,longbibliography,notitlepage]{revtex4}
\usepackage{etex}
\usepackage{amsmath,amssymb,amsthm}
\usepackage[colorlinks=true,citecolor=blue,urlcolor=blue]{hyperref}
\usepackage[pdftex]{graphicx}
\usepackage{times,txfonts}
\usepackage{braket}
\usepackage{color}
\usepackage{natbib}
\usepackage{amsmath,blkarray}
\usepackage{mathtools}
\usepackage{latexsym}
\usepackage{tabularx, booktabs}
\usepackage{graphics,epstopdf}
\usepackage{graphicx}
\usepackage{float}
\usepackage{graphicx}
\usepackage{amsfonts}
\usepackage{subcaption}
\usepackage{color,soul}

\newcommand{\be}{\begin{equation}}
	\newcommand{\ee}{\end{equation}}
\newcommand{\ba}{\begin{eqnarray}}
	\newcommand{\ea}{\end{eqnarray}}

\newtheorem{definition}{Definition}
\newtheorem{proposition}{Proposition}

\newtheorem{theorem}{Theorem}
\newtheorem{corollary}{Corollary}[theorem]

\begin{document}
\title{Constrained Measurement Incompatibility from Generalised Contextuality of Steered Preparation}
\author{ Sumit Mukherjee$^{1}$ }
\email{mukherjeesumit93@gmail.com}
\affiliation{Department of Physical Sciences, Indian Institute of Science Education and Research Kolkata,
Mohanpur 741246, West Bengal, India,}
\affiliation{Korea Research Institute of Standards and Science, Daejeon 34113, South Korea}
\author{ A. K. Pan }
\email{akp@phy.iith.ac.in}
\affiliation{Department of Physics, Indian Institute of Technology Hyderabad, Telengana-502284, India}
\begin{abstract}
   In a bipartite Bell scenario involving two local measurements per party and two outcomes per measurement, the measurement incompatibility in one wing is both necessary and sufficient to reveal the nonlocality. However, such a one-to-one correspondence fails when one of the observers performs more than two measurements. In such a scenario, the measurement incompatibility is necessary but not sufficient to reveal the nonlocality. In this work, within the formalism of general probabilistic theory (GPT), we demonstrate that unlike the nonlocality, the incompatibility of $N$ arbitrary measurements in one wing is both necessary and sufficient for revealing the generalised contextuality for the sub-system in the other wing. Further, we formulate an elegant form of inequality for any GPT that is necessary for $N$-wise compatibility of $N$ arbitrary observables. Moreover, we argue that any theory that violates the proposed inequality  possess a degree of  incompatibility that can be quantified through the amount of violation. We claim that it is the generalised contextuality that provides a restriction to the allowed degree of measurement incompatibility of any viable theory of nature and thereby super-select the quantum theory. Finally, we discuss the geometrical implications of our results.
\end{abstract}
\maketitle

\section{Introduction}
\footnote{Corresponding author}
It is a distinctive feature of quantum theory over classical theories that the joint probability of measurement outcomes of two observables does not exist, in general. However, if two observables are commuting,  they are co-measurable, i.e., the joint probability of them always exits. Therefore, the commutativity is a sufficient condition for the joint measurability of observables. The textbook quantum theory often deals with projective measurements. However, the most general measurement in quantum theory is represented by positive operator-valued measures (POVMs) \cite{busch} corresponding to an observable, and the projective value measure is just a special case. 

A set of POVMs are called jointly measurable or compatible if there exists a single global POVM that can be suitably marginalized to realize each element of the POVMs from the set. Otherwise, the set of POVMs are called incompatible. Thus, the notion of measurement incompatibility is more general than commutativity, i.e., the latter implies the former but converse does not always hold. Note also that for the case of projective value measure, the pairwise commutativity of a set of arbitrary $N$ number of observables implies the global commutativity of all the $N$ observables. In contrast, the pairwise compatibility of arbitrary $N$ POVMs corresponding to $N$ non-commuting observables does not generally imply the global compatibility of the set \cite{kunjwal14,kunjwal14g,liang11,kunjwal20,yu13}. 

There are many different ways to define the POVMs corresponding to a given observable. However, the unsharp \cite{buschemi24,busch} POVMs are commonly discussed ones, which are smeared versions of projective measurements and directly arise from the non-ideal measurement process in quantum theory. Given a set of $N$ unsharp POVMs that are globally incompatible, it can be made compatible if the degree of unsharpness is lower than a critical value \cite{kunjwal14,kunjwal14g}. Moreover, the greater the incompatibility of a set of POVMs, the higher the  unsharp parameter needs to be taken in order to make the set compatible. Thus, the upper bound to the unsharpness parameter that makes the set of POVMs compatible determines the degree of measurement incompatibility possessed by the concerned measurements.

Over the last two decades, a flurry of important studies have been made to explore the connection between measurement incompatibility and various forms of quantum correlations. It is well-known that measurement incompatibility is necessary to witness Bell nonlocality \cite{bell,chsh,brunnerrev}. Moreover, in the bipartite Bell scenario \cite{chsh,brunnerrev} where each of the observer implements only two dichotomic measurements, it is proved \cite{wolf09,banik13} that whenever the measurements performed in one of the wings are incompatible, it is always possible to find shared states and measurements on the other wing  such that the joint probability violates the Clauser-Horne-Shimony-Halt inequality. However, for the case of more than two measurements per wing, the measurement incompatibility does not warrant the revelation of nonlocality \cite{bene18,hirsch18} in general. This qualitatively resembles the fact that for all entangled states, the  Bell nonlocality cannot be demonstrated \cite{hororev,guhnearev,werner89}. Thus, measurement incompatibility is necessary but not sufficient to witness Bell nonlocality. 

In contrast, it has been recently shown \cite{tavakoli20} that a form of non-classical quantum correlation - the preparation contextuality \cite{spekkens05,spekkens21}, can be revealed if and only if the measurements are incompatible. This is closely related to the fact that there exists a one-to-one mapping between measurement incompatibility and a weaker form of nonlocality, the notion of steering \cite{uola15,quintino14}.
Note that, the non-classical correlations, such as nonlocality and steering, are exhibited for  specially separated  parties. On the other hand, generalised contextuality \cite{spekkens05} refers to a form of nonclassicality that can be revealed through statistics that correspond to a single system. 

However, one can leverage generalised contextuality from any of the local systems of a nontrivially correlated bipartite global system by considering that the local measurement on any one of the systems remotely prepares another system in the other spatially separated wing \cite{schmid18}. In particular, for quantum theory, such nontrivially correlated systems are represented as entangled states, which has recently been shown \cite{plavala24} as a prerequisite for obtaining the generalised contextuality of the remotely prepared system. Thus, measurement in one of the wings in a bipartite Bell test can constrict the maximum amount of generalised contextuality in the other wing. In this context, for quantum theory it is already established \cite{uola15} that in the discussed bipartite scenario the local assemblage of a particular wing is unsteerable if and only if the measurements performed on the other wing are compatible. Supplementing this result, it is also found \cite{tavakoli20} that if the measurements on one wing are compatible, then the resultant statistics obtained from the systems on other wing must admit preparation non-contextual model and vice-versa.

Considering the growing acceptability of generalised quantum contextuality as a viable notion of nonclassicality behind a plethora of information processing and computational tasks \cite{schmid18,bae21,mukherjee22,clonning20,spekk09,pan19,saha19,schmid22C}, it is important to investigate what structural uniqueness of quantum theory providing such contextual advantage in those tasks. On the other hand, it is also well known that incompatible quantum measurements implies advantage in a number of tasks including quantum state discrimination \cite{Carmeli19,skrzypczyk19}, quantum random access codes \cite{carmeli20,saha23}, and parameter estimation \cite{jae23}. This indicates that  there must be some interrelation between measurement incompatibility and generalised contextuality that are not only significant for quantum theory but also for any other operational theory. Therefore, possible existence of such an interconnection is essential for comprehending the intricacies of quantum theory that singles it out from any other possible no-signalling operational theories. 
%In this paper, we try to take a step forward in that direction.}

Against this backdrop, in this work,  we investigate this issue from the perspective of generalised probabilistic theory (GPT) \cite{plavala21, janotta11, janotta14, barrett07, hardy01}. We consider a bipartite Bell scenario involving two parties Alice and Bob, who share a non-trivially correlated system. Alice randomly performs measurements of an arbitrary $N$  number of dichotomic observables in her local subsystem, and Bob randomly performs a certain number of measurements on his part of the system and collects the joint statistics. The primary aim of this paper is to show that in any no-signalling theory, if Alice's measurements are compatible, then any statistics obtained from Bob's subsystem admit the generalised non-contextual model. %Obviously, in order to reveal the existence of generalised contextuality of Bob's subsystem he has to implement suitable measurements on his subsystem. 
In particular, we argue that, unlike the case of nonlocality, for any no-signalling operational theory, the measurement incompatibility of an arbitrary $N$ number of dichotomic observables in one wing is both necessary and sufficient for revealing the generalised contextuality for the steered subsystems of the other wing. In particular, by invoking the framework of GPT, we demonstrate that the degree of generalised contextuality in a GPT determines the maximum allowed incompatibility that the theory possesses. Furthermore, by noting the geometrical meaning \cite{spekkens21} of generalised contextuality we connect the compatibility of measurements at one wing with the geometrical structure of the GPT describing the local statistics of other wing. More precisely, we prove that the local statistics of the steered preparations of Bob admits a \textit{simplex-embeddable } GPT description if and only if Alice's measurements are compatible.

 Further, in order to establish our key results, we introduce a measure of incompatibility in a GPT by incorporating the fact that the optimal value of fuzziness for which an arbitrary $N$ number of POVMs becomes compatible depends largely on the structure of the no-signalling theory. In this regard, witnesses for pairwise incompatibility of observables in the framework of GPT has already been derived in \cite{beneduci22,Filippov17}. Here, we construct an inequality based witness for $N$-wise compatibility valid for any no-signalling probabilistic theory. This witness enables us to quantify the amount of incompatibility of an arbitrary $N$ number of observables corresponding to a particular theory. Having this tools in hand we finally argue that if quantum theory is taken as the governing algorithm of nature then a class of theory whose optimal incompatibility is more or less than quantum theory must not exist. To put it more precisely, by revealing a hitherto unexplored link between two indispensable notions of classicality, namely, the compatibility and the generalised noncontextuality, our result quantifies the maximum degree of measurement incompatibility that nature possesses.

This paper is organized as follows. In Sec. II. we discuss the operational theories and ontological models to finally discuss the notion of classicality. In Sec. III we describe the GPT and its connection to ontological models. In Sec. IV we define the incompatibility of measurements in a GPT and discuss the incompatibility of a theory. Sec. V contains our main arguments and results. In Sec. VI we discuss the optimal incompatibility of measurements allowed in quantum theory while in Sec. VII. we sketch some geometrical implications of our results and finally conclude with future plans in Sec. VIII.

\section{Ontological model of an operational theory and classicality}
\label{Sec2}
Before proceeding to discuss how the notion of classicality is described in terms of the entities of an operational theory we  briefly encapsulate the notion of an operational theory. The self-explanatory term 'operational' includes all the possible operable entities involved in an experiment. The sole aim of an operational theory is to define these entities to predict the statistics that are obtained in any given experiment. However, a reliable operational theory not only perfectly predicts the experimental statistics obtained at present but should also be able to determine any statistics that can be obtained in future experiments. 

%Conventionally, such theories consist of three basic building blocks, namely, a set of preparation procedures, a set of transformation procedures, and a set of measurement procedures. In addition to this, the theory defines a rule for obtaining the statistics of an experiment in terms of the preparation, transformation, and measurements. 

Technically speaking, the primitive of an operational theory essentially consists of a set of preparation procedures $\{\mathbb{P}\}$, a set of transformations $\{\mathbb{T}\}$, and measurement procedures $\{\mathbb{M}\}$. However, in this paper, we do not explicitly use transformation procedures. Now, given a preparation procedure $P\in \mathbb{P}$ and a measurement procedure $M\in \mathbb{M}$, an operational theory predicts the statistics of the experiment by providing the probability $p(m|P, M)$ of obtaining a particular output $m$. If the operational theory is quantum theory then the preparation procedure is represented by the density matrix $\rho$ and the measurement procedures are in general represented by a set of POVMs $\{E_{m}\}$ where $m$ is the outcome of the measurement. The Born rule provides the probability of obtaining the outcome $m$ of an experiment so that $p(m|P, M)=Tr[\rho E_{m}]$. Although operational theory perfectly predicts the experimental statistics, in general it does not provide an any causal description of reality behind the statistical predictions. The aim of an ontological model is to provide such a causal description of operational statistics.  %We will see in the following subsections how based on certain abstractions of an operational theory one can define different notions of classicality.

%\subsection{Ontological model of an operational theory and generalised noncontextuality}

An ontological model of an operational theory can succinctly be described as follows \cite{harrigen,spekkens05}. In such a model, it is assumed that a preparation procedure $P$ produces the ontic state having probability distribution $\mu_{P}(\lambda)$ which satisfies, $\int_{\Lambda}\mu_{P}(\lambda) d\lambda =1$. Here, $\lambda \in \Lambda$,  with $\Lambda$ denotes the ontic state space. Given a measurement procedure $M$, each ontic state $\lambda$ assigns a response function $\xi_{m|M}(\lambda)$ corresponding to a particular outcome $m$ of the measurement, with $\sum\limits_{m}\xi_{m|M}(\lambda)=1$, $\forall\lambda$. Note that a viable ontological model of an operational theory has to reproduce all possible statistics that the operational theory predicts. %Physically, the response functions $\xi_{m|M}(\lambda)$ are referred to as the probability of obtaining the outcome $m$ when the preparation corresponds to the ontic state variable $\lambda$. 
As an example, an ontological model of quantum theory must reproduce the Born rule so that
	\begin{equation}
		Tr(\rho E_{m})=\int_{\Lambda}\mu_{P}(\lambda)  \xi_{m|M}(\lambda) d\lambda . \label{eq:born}
	\end{equation}
	
It is important to note here that, in general, any specific entity of the ontological model can be obtained by a one-to-one mapping from the elements of the operational theory for which the ontological model is defined. For example, every preparation procedure of an operational theory is equivalently mapped to a probability distribution over the ontic state variables in an ontological model. In subsequent sections, we argue that this is not always the case when one considers ontological models of other abstractions of an operational theory, for example, the ontological models corresponding to some GPTs.

It has been shown in the literature that the predictions of quantum theory are in conflict with classical models. This necessitates the articulation of the precise notion of classicality in an ontological model of the operational theory. Certainly, one may impose various assumptions on an ontological model which can lead different forms of classicality. Here, we consider a widely accepted notion of classicality- the generalised non-contextuality \cite{spekkens05} .  

The notion of generalised non-contextuality can be made precise by introducing equivalence classes of experimental procedures.  Let two preparation procedures (measurement procedures) $P_{1}$ and $P_{2}$ ($M_{1}$ and $M_{2}$) cannot be operationally distinguished by any measurement (preparation) for all given outcomes $m$. We call these two preparation procedures to be operationally equivalent, i.e.,  $p(m|P_{1}, M)=p(m|P_{2},M) $, $\forall M,m$. Similarly, two operationally equivalent measurement procedures satisfy $p(m|P, M_{1})=p(m|P,M_{2}) $, $\forall P,m$. In quantum theory, such equivalence is generally represented by preparation procedures that produce the same density matrix $\rho$ and measurement procedures that realize the same POVM element $E_{m}$. For explicit examples of states and measurements that are operationally equivalent in quantum theory, we recommend consulting sections IV and V of the ref. \cite{spekkens05}.
 
%Even though the ontological model we are seeking reproduces the Born rule, it does not guarantee that certain restrictions to the operational level will also reflect in the respective probability distributions of ontic state variables $\lambda \in \Lambda$. 
The key question is how such operational equivalences of preparation and measurement procedures in an operational theory can be represented in the ontological model of that operational theory. One may surmise that an equivalent representation of such operational equivalence may also hold in the ontological model. This motivates one to define the notion of generalised non-contextuality \cite{spekkens05} which is essentially described in terms of two basic assumptions imposed in any ontological model defined as follows. If any two preparations $P_{1}$ and $P_{2}$ are operationally equivalent then they are represented by the same ontic state probability distributions, i.e.,   $p(m|P_{1}, M)=p(m|P_{2},M)	\Rightarrow \mu_{P_{1}}(\lambda)=\mu_{P_{2}}(\lambda)$, $\forall m, M$. This constitutes the assumption of preparation non-contextuality. Along the same line, if two operationally equivalent measurements $M_{1}$ and $M_{2}$ are represented by equivalent response functions in the ontological model, then the model is called measurement non-contextual. Technically, in such a model $\forall m, P, \ p(m|P, M_{1})=p(m|P,M_{2})	\Rightarrow \xi_{m|M_{1}}(\lambda)=\xi_{m|M_{2}}(\lambda)$. Simply, a model satisfies both preparation and measurement noncontextual assumptions is called a generalised noncontextual ontological model. The crucial question is up to what extent such a generalised noncontextual ontological model reproduces the statistics of an operational theory. It is already known that certain quantum statistics cannot be reproduced by a generalised noncontextual model., i.e., an ontological model necessarily needs  some amount of generalised contextuality in order to compatible with the operational statistics. In this paper, we consider generalised noncontextuality as the notion of classicality to demonstrate our main results in GPT framework.

\section{GPT and its ontological model}
The idea of GPT \cite{janotta11,janotta14,plavala21,hardy01,barrett07} paves an elegant route towards developing the very framework of possible theories (of the nature) to describe certain observational phenomenon solely from the probabilities. The motivation behind studying the GPTs essentially involves singling out the viable physical theories from a large class of possible theories and pinpointing the underlying reasoning through which one may discard a specific set of  among them.  

In a GPT, the state of a system is associated with element $\omega \in \Omega$, where $\Omega$ forms a compact convex set residing inside the affine span of $\Omega$ defined as \textbf{Aff}$[\Omega]$. The convexity of state space captures the fact that
preparing different systems represented by states $\omega_i$, with probability $p_{i}$, results in the mixture of states $\omega = \sum_{i}p_{i}\omega_{i}$ (convex
combination of $\omega_{i}$), and such mixed states that are allowed in the theory reside in the convex hull of the space $\Omega$. The extremal points of $\Omega$ are the pure states that cannot be written as the convex combination of other states. Without any loss of generality, we assume that \textbf{Aff}$[\Omega]$ in which the state space resides is embedded into a real vectors space $\mathcal{V}$ which additionally possesses inner products defined as $\langle \ , \ \rangle$. For example, in quantum theory, this vector space is a Banach space $\mathcal{B}(\mathcal{H})$ of all linear Hermitian operators from the Hilbert space $\mathcal{H}$ and the inner product is the usual Hilbert-Schmidt inner product $\langle \ , \ \rangle_{HS}$.

 In addition to the states, a GPT  contains entities that are representative of measurements. For instance, an outcome $m$ of measurement $M$ is represented through the so-called effect
$e_{\tiny M}^{\tiny m} \in \Xi$ where $\Xi$ is the effect space. The effect $e_{\tiny M}^{\tiny m}$ is a linear map from a state $\omega$ to a valid probability  $p(m|\omega,M)=\langle e_{\tiny M}^{\tiny m}, \omega \rangle$ of measurement outcome, i.e., $e_{\tiny M}^{\tiny m} : \Omega \rightarrow [0,1]$ , such that $\forall \omega \ $  $\sum_{m}\langle e_{\tiny M}^{\tiny m},\omega\rangle = 1$. Moreover, there exists a unit measure $\textit{u}$ for the unique effect $\textit{u}$ : $\Omega \rightarrow 1$  that corresponds to a measurement having only one outcome that is certain to occur given any state $\omega$ from the state space. Similar to the embedding of the state space in a vector space $\mathcal{V}$ we will assume that the effect space $\Xi$ will be embedded into the dual of the mentioned vector space defined as $\mathcal{V}^*$. Such an embedding is convenient in the sense that one can represent both the state and the effect spaces as hyperplanes and volumes, respectively, embedded into the vector spaces of same dimension. Any valid probability distribution  $\langle e_{\tiny M}^{\tiny m}, \omega \rangle = p(m|\omega,M)$ is then a bilinear map $\mathcal{M}$: $\mathcal{V} \times \mathcal{V}^* \rightarrow \mathbb{R}$.

In general, a GPT describing single system's states and effects must have the aforementioned properties. However, specific GPTs can be formulated by imposing additional restrictions, which are termed fragments \cite{selby23} of GPTs. In this paper, we restrict ourselves to this general definition of GPTs and do not assume that every logically possible states and effects are physically realisable- an assumption commonly  known as the \textit{no-restriction hypothesis} \cite{janotta13}. 

\subsection{Description of composite system in GPT}
So far, we have considered the GPTs corresponding to a single system only. For our purpose here, it is essential to discuss some important properties of GPTs that involve composite systems, in particular GPTs that involve bipartite systems.  Let us consider the bipartite scenario involving two distant parties (Alice and Bob) who are in possession of their individual local states $\omega_{A}^{i}$ and $\omega_{B}^{j}$ prepared in the i$^{th}$ and j$^{th}$ run of the experiment respectively. The simplest form of the joint system can be described by the convex mixture of the tensor products of local states as, $\omega_{AB}=\sum_{i,j}c_{ij}\omega_{A}^{i} \otimes \omega_{B}^{j}$ with $\sum_{i,j}c_{ij}=1$ . Similarly, the joint effect corresponding to the measurements performed by the parties can also be written as, $e_{AB}=\sum_{i,j} d_{ij}e_{A}^{i} \otimes e_{B}^{j}$ with $\sum_{i,j}d_{ij}=1$, where $e_{A}^{i}$ and $e_{B}^{j}$ are the effects corresponding to the local measurements. Below, we formally define the mathematical operation viable for such states and effects corresponding to a GPT.

\begin{definition}{Minimal tesor products:  In a GPT framework the viable mathematical operation for which every set of bipartite states and effects can be written in the following form,

\begin{eqnarray}\label{minten}
    &&(\mathcal{V}_{A}\otimes_{min}\mathcal{V}_{B}) \nonumber \\
    &&=\Big\{ \omega_{AB}=\sum_{i,j}c_{ij}\omega_{A}^{i} \otimes \omega_{B}^{j} \ \big| \omega_{A}^{i}\in \mathcal{V}_{A},  \omega_{B}^{j}\in \mathcal{V}_{B}, 0 \leq c_{ij} \leq 1 \Big\} \nonumber \\
&&(\mathcal{V}^{*}_{A}\otimes_{min}\mathcal{V}^{*}_{B}) \nonumber \\
    &&=\Big\{ e_{AB}=\sum_{i,j}d_{ij}e_{A}^{i} \otimes e_{B}^{j} \ \big| e_{A}^{i}\in \mathcal{V}_{A}^{*},  e_{B}^{j}\in \mathcal{V}_{B}^{*}, 0 \leq d_{ij} \leq 1 \Big\} 
\end{eqnarray}} \label{simplicialdef}

is known as minimal tensor products. 
\end{definition}
Thus, any extremal joint state on the global state space $\mathcal{V}_{A}\otimes_{min}\mathcal{V}_{B}$ of bipartite systems is just the product of states of local subsystems. However, this is not the only way to write the global states or the joint effects. For example if the states or effects of these bi-partition are correlated then they can not be written as the convex combination of product states and effects respectively. When the extremal states on global state space cannot be written as products of local states it is represented by another mathematical operation complementary to the minimal tensor products and are generally defined through what is called maximal tensor products defined as follows.

\begin{definition}{Maximal tesor products: \label{simplicialdef}
In a GPT framework the mathematical operation that represents the maximal set of bipartite states and effects is called the maximal tensor products and is defined as

\begin{eqnarray}\label{maxten}
\label{simplicialdef}
    &&(\mathcal{V}_{A}\otimes_{max}\mathcal{V}_{B}) \nonumber \\
    &&=\Big\{ \omega_{AB} \in \mathcal{V}_{A} \otimes \mathcal{V}_{B} \ \big| \langle e_{A} \otimes e_{B},\omega_{AB} \rangle \geq 0 , \forall e_{A} \in \mathcal{V}_{A}^{*}, e_{B} \in \mathcal{V}_{B}^{*}\Big\} \nonumber \\
    &&(\mathcal{V}^{*}_{A}\otimes_{max}\mathcal{V}^{*}_{B}) \nonumber \\
    &&=\Big\{ e_{AB} \in \mathcal{V}_{A}^{*} \otimes \mathcal{V}_{B}^{*} \ \big| \langle \omega_{A}\otimes \omega_{B} e_{AB} \rangle \geq 0 , \forall \omega_{A} \in \mathcal{V}_{A}, \omega_{B} \in \mathcal{V}_{B}\Big\} \nonumber \\
\end{eqnarray}}

\end{definition}

Strictly speaking, the maximal tensor product states are the largest set of bipartite states that return valid probabilities when the inner product is taken with any effect from minimal tensor products. This definition is the same for maximal tensor product effects, only the role of states and effects interchanges. In this paper, we will use the term \textit{nontrivially correlated states} to refer to the set of states that can not be written as the states belonging to minimal tensor products. In quantum theory, such states are generally called entangled states. However, in the GPT framework, we prefer to use the term nontrivially correlated states that includes the entangled states too.

\subsection{Ontological model of a GPT and generalised noncontextuality}
\label{OntGPT}

We have already discussed the ontological model of an operational theory. Now, we consider the ontological model of a GPT. More precisely, we try to find the answer to the question whether there exists an ontological model of a GPT too. For this, we first recall the correspondence between the entities in an operational theory and a GPT in a more technical manner. Consider a specific preparation procedure $P_{\epsilon} \in \mathbb{P}$ and represent the equivalent class of preparation procedures of $P_{\epsilon}$ as  $[P_{\epsilon}]_{\equiv} \doteq \{P_{\epsilon'} \ | \ P_{\epsilon} \equiv P_{\epsilon'}\}$. Here, we use $`\equiv'$ to represent the equivalent relations on the set of all preparations $\mathbb{P}$. Then, the quotient of $\mathbb{P}$ by $\equiv$ is the set $\mathbb{P}/\equiv \ \doteq \{[P_{\epsilon}] \ | \ P_{\epsilon} \in \mathbb{P}\}$. Similarly, with the equivalent relations $\equiv$ on the set $\mathbb{M}$ of measurements let the equivalence class of an element ${M_{\epsilon}} \in \mathbb{M}$ is the set $[M_{\epsilon}]_{\equiv} \doteq \{M_{\epsilon'} \ | \ M_{\epsilon} \equiv M_{\epsilon'}\}$ and the quotient of the set $\mathbb{M}$ by $\equiv$ is the set $\mathbb{M}/\equiv \ \doteq \{[M_{\epsilon}] \ | \ M_{\epsilon} \in \mathbb{M}\}$. Then the GPT states and effects can be represented by the following maps as, 

\begin{equation}
\label{cri1}
    \omega_{P}: \mathbb{P} \rightarrow \Omega \ \ \text{and} \ \  e_{M}^{m}: \mathbb{M} \rightarrow \Xi
\end{equation}
 with $\Omega$ $(\Xi)$ forms the set of GPT states (effects) isomorphic with the set $\mathbb{P}/\equiv$ $(\mathbb{M}/\equiv)$ of the operational theory.

 With the aforementioned representation it is now evident that the equivalence class of preparations and measurements are represented by \emph{the  same} state and effect in the GPT respectively. Thus, an ontological model defined for a particular GPT exists if it can accommodate states as maps from GPT state space $\Omega$ to probability distributions on ontic state space $\Lambda$, and the effects as maps from the GPT effect space $\Xi$ to another valid probability distribution over $\Lambda$. This can be formally written as

\begin{equation}
\label{cri2}
    \bar{\mu}_{\omega_P}: \Omega \rightarrow \mathcal{L}(\Lambda) \ \ and \ \  \bar{\xi}_{e_{M}^{m}}: \Xi \rightarrow \mathcal{L}(\Lambda)
\end{equation}
 where the maps $\bar{\mu}_{\omega_P}$ and $\bar{\xi}_{e_{M}^{m}}$ represent the ontic state and response function of the ontological model of the GPT corresponding to the GPT state $\omega_P$ and effect $e_{M}^{m}$ respectively.
 
At this point, we recall an important result from Ref.  \cite{spekkens21} that will be used to establish our main result. We encapsulate that result in the following theorem and provide a straightforward proof of it.

\begin{theorem}{}
\label{spekktheor}
There exists an ontological model of the GPT corresponding to a specific operational theory if and only if the ontological model of the operational theory is generalised noncontextual.
\end{theorem}

\textit{Proof:} We need to demonstrate that if the ontological model of an operational theory is generalised noncontextual then the GPT corresponding to the operational theory must have an ontological model and vice-versa. The assumption of preparation noncontextuality dictates the following.

\begin{eqnarray}
    P_{1} \equiv P_{2} &&\implies \mu_{P_{1}} (\lambda)= \mu_{P_{2}} (\lambda) \\
 &&\implies \sum_{\lambda} \xi_{m|M}(\lambda)\mu_{P_{1}}(\lambda) =\sum_{\lambda} \xi_{m|M}(\lambda)\mu_{P_{2}}(\lambda) \ \ \forall m, M \nonumber
 \end{eqnarray}
 This implies that in terms of the entities of operational theory the following holds,
 \begin{eqnarray}
 \label{gptO}
 && P(m|M,P_{1}) = P(m|M,P_{2}) \ \ \forall m, M \nonumber \\
 &&\implies \langle \omega_{P_{1}}, e_{M,\omega}^{m} \rangle = \langle \omega_{P_{2}}, 
e_{M,\omega}^{m}\rangle \ \ \ \forall e_{M,\omega}^{m}
\end{eqnarray}

Now, for the ontological model corresponding to the GPT, Eq.\eqref{gptO} implies the following.

\begin{eqnarray}
&& \sum_{\lambda} \bar{\xi}_{e_{M}^{m}}(\lambda)\bar{\mu}_{\omega_{P_{1}}}(\lambda) =\sum_{\lambda} \bar{\xi}_{e_{M}^{m}}(\lambda)\bar{\mu}_{\omega_{P_{2}}}(\lambda) \ \ \ \ \forall \bar{\xi}_{e_{M}^{m}}(\lambda) \nonumber \\ \nonumber \\
&&\implies  \bar{\mu}_{\omega_{P_{1}}}(\lambda) = \bar{\mu}_{\omega_{P_{2}}}(\lambda) \nonumber
\end{eqnarray}

This proves the existence of the ontic states for the ontological model corresponding to the GPT that the operational theory defines. The converse implication is quite straightforward as follows.
 \begin{eqnarray}
 P_{1} \equiv P_{2} &\implies&  \bar{\mu}_{\omega_{P_{1}}}(\lambda) = \bar{\mu}_{\omega_{P_{2}}}(\lambda) \nonumber \\
 &\implies& \mu_{P_{1}} (\lambda)= \mu_{P_{2}} (\lambda)
 \end{eqnarray}
 Following the similar line of argument, we can run the proof for effects, so that
 \begin{eqnarray}
     {M}_{1} \equiv {M}_{2} &\implies & \xi_{m|M_{1}}(\lambda) = \xi_{m|M_{2}}(\lambda) \ \forall m, \lambda \\
     &\iff& \bar{\xi}_{e_{M_{1}}^{m}}(\lambda)= \bar{\xi}_{e_{M_{2}}^{m}}(\lambda)  \ \forall m, \lambda 
 \end{eqnarray}

This completes the proof.

\section{Notion of measurement incompatibility in a GPT}

Given the state space $\Omega$,  any measurement in the GPT formalism is represented by an affine map $M: \Omega \rightarrow \mathcal{P}(\Omega) $, where $\mathcal{P}(\Omega)$ constitutes the set of probabilities for the outcomes of the measurement. Technically, a measurement $M$ on a system  state $\omega \in \Omega$ producinges dichotomic outcomes $m \in \{\pm1\}$  can be represented as, $M=\{e_{\tiny M,\omega}^{\tiny m}| \sum_{m}e_{\tiny M,\omega}^{\tiny m}=1, e_{\tiny M,\omega}^{\tiny m}\geq 0 ,\forall \omega\in \Omega\}$. Therefore,  the expected average value of $M$ can be written as

\begin{equation}
    \langle M \rangle = e_{\tiny M,\omega}^{\tiny +1} - e_{\tiny M,\omega}^{\tiny -1}
\end{equation}
Now, two dichotomic observables $M_{1}$ and $M_{2}$ in a GPT is said to be compatible if there exists joint probability distribution for its outcomes ($m_1$ and $m_2$) that can be derived from a joint effect $e_{\tiny M_{1},M_{2},\omega}^{\tiny m_{1},m_{2}}$ corresponding to the concerned observables such that upon suitable marginalization it returns the individual effects. This is represented as follows,

\begin{eqnarray}
   \forall m_{2}, \ \ \sum_{m_{1}} e_{\tiny M_{1},M_{2},\omega}^{\tiny m_{1},m_{2}} &=& e_{\tiny M_{2},\omega}^{\tiny m_{2}}   \nonumber\\
       \forall m_{1}, \ \ \sum_{m_{2}} e_{\tiny M_{1},M_{2},\omega}^{\tiny m_{1},m_{2}} &=& e_{\tiny M_{1},\omega}^{\tiny m_{1}}
\end{eqnarray}

Such joint effects corresponding to a pair of observables may not always exist when only perfect measurements are considered. However, for the imperfect or fuzzy measurement of a pair of observables,  such a joint effect may exist if the degree of fuzziness is upper bounded by a certain critical value \cite{busch13}. We call such fuzzy version of the measurements as unsharp measurement and represent the unsharpness parameter as $\eta\in[0,1]$. %A pair of observables is called maximally incompatible when for this pair $\eta = \eta_{max}$, where $\eta_{max}$ is the largest value of $\eta$ above which no pair of observables are jointly measurable. 

The notion of compatibility can be further extended to any arbitrary number of observables. A set of $N$ observables $\{M_{i}\}_{i=1}^{N}$ are said to be  compatible \textit{iff} there exists a joint effect \begin{eqnarray}\{e_{\tiny M_{1},M_{2}...M_{N},\omega}^{\tiny m_{1},m_{2}...m_{N}}|\sum_{m_{1},m_{2}...m_{N} \in \{\pm 1\}}e_{\tiny M_{1},M_{2}...M_{N},\omega}^{\tiny m_{1},m_{2}...m_{N}}=1\}
\end{eqnarray}
such that, the suitable marginalization of the joint effect returns the single  effect, i.e.,

\begin{equation}
\label{nwise}
    \forall m_{N}, \sum_{m / m_{N}} e_{\tiny M_{1},M_{2}...M_{N},\omega}^{\tiny m_{1},m_{2}...m_{N}}= e_{\tiny M_{N},\omega}^{\tiny m_{N}}
\end{equation}

 If a set of $N$ observables is $N$-wise compatible then they are certainly $(N-l)$-wise compatible for all $l< N$. However, in general, the converse is not true. In the following, we define a particular measure of incompatibility of a GPT which  we call  $N^{\eta^{*}}$-compatible GPT to quantify the degree of $N$-wise compatibility of observables corresponding to that GPT. We introduce the following definition, which will be used to establish our main arguments.

\begin{definition}{$N^{\eta^{*}}$-compatible GPT:} \label{defeta}
A specific GPT is said to be $N^{\eta^{*}}$-compatible GPT if the necessary condition for the $N$-wise compatibility of the fuzzy measurement of $N$ arbitrary mutually-maximally incompatible observables of that GPT is $\eta\leq\eta^{*}$.
\end{definition}

Note that for any given value of $N$, the higher the value of $\eta^{*}$, the less $N$-incompatible the GPT is. For instance, the extreme case of $\eta^{*}=1$ indicates that the GPT is $N-$wise compatible, even if the measurements are perfect. Such GPTs  are deemed classical in terms of incompatibility, as those $N$ observables have simultaneous values. Following definition \ref{defeta}, a GPT can be said to be maximally incompatible if in such a GPT, the highest amount of unsharpness $(\eta^{\ast})$ needs to be given to make any arbitrary $N$ observable compatible. It is shown in \cite{heinosaari16} that if $\eta^{*}=\frac{1}{N}$ then all $N$ observables are compatible in any arbitrary GPT.  Therefore, a GPT is referred to  as maximally incompatible if $\eta^{*}=\frac{1}{N}$ \cite{heinosaari16}. However, it may be possible that in a specific GPT for $\eta >\frac{1}{N}$ is enough to establish $N$-wise compatibility.

We note here that in quantum theory, the maximally incompatible pair of observables do not exist \cite{heinosaari16,heinosaari14} in finite-dimensional Hilbert space. However, if the dimension of the quantum system is infinite, one needs the unsharpness parameter $\eta=1/2$ to make an arbitrary pair of observables compatible. On the other hand, any pair of qubit observables becomes compatible if and only if the unsharpness parameter is $\eta \leq 1/\sqrt{2}$ \cite{busch}. Later, this result is generalised \cite{banik13}  for any pair of dichotomic observables in an arbitrary finite dimension of Hilbert space. We remark here  that a very few results regarding $N$-wise compatibility in any arbitrary dimension $d$ are known.  The best known \cite{wiseman07} tight bound on the unsharpness for the compatibility of set of $N$ noisy measurements is $\eta \leq \frac{H_{d}-1}{d-1}$, where , $H_{d}=\sum_{n=1}^{d}\frac{1}{n}$ is called the $d^{th}$ harmonics. It has also been argued in \cite{wiseman07,quintino14,uola14} that for any dimension $d$, the above bound is necessary and sufficient for the compatibility of arbitrary number of measurements. For example, an arbitrary $N$ noisy qubit measurements becomes compatible iff $\eta\leq 1/2$ \cite{quintino14}. By noting the connection between compatibility of a set of measurements with quantum steering, this bound is also supported by recent results \cite{chitambar24,renner24}.

Given the above perspective, we ask the more general question of what is the necessary condition for $N$-wise compatibility for any no-signalling GPT. From the quantum mechanical perspective, such a $N$-wise compatibility condition enables us to find the critical value of $\eta$ required to make an arbitrary $N$ number of dichotomic observables compatible.  %However, we could only derive a necessary condition and using that condition we derived the necessary condition for $N$-wise compatibility in terms of the number $N$ for quantum measurements represented by operators living in a unknown dimension.

\section{Landscape of the arguments and Results}
In the preceding sections, we laid out the essential prerequisites imperative for establishing our arguments. Now, we embark on a detailed exploration of these prerequisites to convey the key findings of this paper. In the following, we present the main result in two installments. First, we establish the correspondence between measurement compatibility and generalised contextuality, and second, we deduce a necessary condition for measurement compatibility itself.
%\subsection{Correspondence between measurement compatibility and generalised non-contextuality}

We consider the following bipartite scenario that features two spatially separated parties (Alice and Bob) who share  a state $\Omega_{AB}$. The party Alice (Bob) performs measurements of $N$ ($2^{N-1}$) observables $\{A_{x}\}_{x=1}^{N}$ ( $\{B_{y}\}_{y=1}^{2N-1})$ on her (his) local subsystem. If the shared state $\Omega_{AB}$ is non-trivially correlated then Alice's measurements can remotely prepare the states in Bob's wing.  Without loss of generality, we can assume that the composite system is represented as a state that cannot be written in a product form (minimal tensor product states)  given in Eq. \eqref{minten}. However, we also do not intend to claim that the state is of other extremal class (maximal tensor product states) given by Eq. \eqref{maxten}. It just suffices to consider any state $\Omega_{AB}$ entailing nontrivial global correlations that cannot be obtained from separable states. In such a scenario the following proposition and its proof is worth mentioning before discussing our main results.

\begin{proposition}{}
\label{lambdasteer}
 Any local statistics produced by the remotely prepared states at Bob's wing can be modelled by states associated with an ontic variable $\lambda$, if and only if Alice's measurements are compatible. 
\end{proposition}

\textit{Proof:} As defined, Alice randomly performs the measurements of dichotomic observables from the set $\{A_{x}\}_{x=1}^{N}$ on her wing. In terms of the GPT effects, these observables can be  represented as

\begin{equation}
\label{eq:jmm}
    A_{x}= \Big\{ e^{a_{x}}_{A} \ | \ e^{a_{x}}_{A} \geq 0; \sum_{a_{x}=\{\pm 1\}} e^{a_{x}}_{A} = u_{A}\Big\}
\end{equation}

Alice's specific measurement $A_{x}$  remotely prepares a (un-normalized) state at Bob's wing can be written as,

\begin{equation}
\omega_{B}^{a_{x}}= [ \big(e^{a_{x}}_{A}\otimes \textit{u}_{B}\big), \omega_{AB}]
\end{equation}
where $a_x\in\{+1,-1\}$ is the measurement outcome. It is to be noted here that the mathematical operation $[,]$ or the rule with which the conditional state is obtained in not relevant here. The rule can be fixed for a particular theory of interest. For example, in quantum theory, the marginal states can be obtained by taking partial trace on the composite system upon applying the measurement operator.  

If Alice's measurements are $N$-wise compatible then there exists a global effect $e^{a}_{A}$, with $a=\{a_{x}\}_{x=1}^{N}$ such that all the single effects can be obtained by suitably marginalizing the global effect as

\begin{equation} 
\label{eq:JM}
     e^{a_{x}}_{A}= \sum_{a/ a_{x}} e^{a}_{A}, 
\end{equation}
with $e^{a}_{A}>0$ and $\sum_{a} e^{a}_{A}= u_{A}$ is the unit effect on Alice's subsystem. Thus, Alice's observables are compatible in the sense of Eq. \eqref{eq:JM}, then the remotely prepared states at Bob's wing is given by,

\begin{equation}
\label{eq:steerstate}
\omega_{B}^{a_{x}}= [ \big(\sum_{a/ a_{x}} e^{a}_{A}\otimes \textit{u}_{B}\big), \omega_{AB} ] 
\end{equation}
At this point, we argue that an ontological model corresponding to the remotely prepared state exists if one can assign a set of ontic variables $\lambda\in \Lambda$, such that upon the specification of that set, all the states in Eq. \eqref{eq:steerstate} can be determined. This then means that if the decomposition of the state in Eq. \eqref{eq:steerstate} in terms of $\lambda$ exists, then it can always be written in the following form,

\begin{equation}
\label{eq:steer}
    \omega_{B}^{a_{x}}= \sum_{\lambda \in \Lambda} \ p(\lambda) p(a_{x}|A_{x},\lambda) \ \omega^{\lambda}
\end{equation}
where $\omega^{\lambda}$ is the GPT state corresponding to variable $\lambda$. Non-existence of such decomposition of the states is usually referred to as steering \cite{schrodinger,wiseman07} in literature. However, we do not intend to discuss the very details of steering. Instead, we  use the proof for the existence of states in the form Eq. \eqref{eq:steer} as a helping tool for proving one of our main results articulated in Theorem 2, appearing later in this section. 

To show that a decomposition as Eq. \eqref{eq:steer} exists for the reduced states at Bob's wing, let us first assume that there is a one-to-one correspondence between the elements of $\Lambda$ and $a$ such that $\lambda= a_{x}$ specifies one of the remotely prepared states. The variable $\lambda$ is distributed according to the probability distribution $p(\lambda)= \langle e^{a_{x}}_{A},\omega_{A}^{u_{B}}\rangle$, where $\omega_{A}^{u_{B}}$ is Alice's part of the state when Bob measures the unit effect $u_{B}$ on his part of the subsystem. In each run of experiment Alice sends her local state $ \omega^{a_{x}}_{B}=[ (e^{a_{x}}\otimes u_{B}), \omega_{AB} ]$ to Bob. Furthermore, when Bob asks her to perform a specific measurement, say $x$, and announce the outcome, she simply conveys the outcome $a_{x}$ according to the probability distribution $p(a_{x}|x,\lambda)=\delta_{\lambda,a_{x}}$. Thus each of the states that Bob gets can be written as Eq. \eqref{eq:steer}. The correspondence between the existence of such decomposition of remotely prepared states and the joint measurability has also been demonstrated in ref. \cite{jencova22, manik15j}.

Now, we prove the converse. Let us assume that the decomposition of the form Eq. \eqref{eq:steer} exists, i.e., 
\begin{equation}
\label{eq:unsteer}
    \langle(e^{a_{x}} \otimes u_{B}),\omega_{AB} \rangle = \sum_{\lambda \in \Lambda} p(\lambda) \ p(a_{x}|x,\lambda) \ \omega^{\lambda}
\end{equation}

Since $\omega^{\lambda}$ is the local state of Bob remotely prepared due to Alice's measurement result corresponding to the effect $e^{\lambda}$, it can be written as,

\begin{equation}
\label{dd}
    \omega^{\lambda} = \langle(e^{\lambda} \otimes u_{B}),\omega_{AB} \rangle 
\end{equation}

Using Eq. (\ref{dd}), the Eq. \eqref{eq:unsteer} can be re-written as,

\begin{equation}
\label{eq:unsteer1}
    \langle(e^{a_{x}} \otimes u_{B}),\omega_{AB} \rangle = \sum_{\lambda \in \Lambda} p(\lambda) \ p(a_{x}|x,\lambda) \ \langle(e^{\lambda} \otimes u_{B}),\omega_{AB} \rangle
\end{equation}
    
Comparing the left and right hand side we can easily see that $e^{a_{x}}_{A}$ in terms of a global effect $e^{\lambda}_{A}$ as,

\begin{equation}
\label{eq:gpovm}
    e^{a_{x}}_{A}= \sum_{\lambda \in \Lambda} p(\lambda) \ p(a_{x}|x,\lambda) \ e^{\lambda}_{A} 
\end{equation}

By taking $\Lambda=a$ and $p(\lambda) \ p(a_{x}|x,\lambda)=1-\delta_{a_{x},a}$ we finally recover Eq. \eqref{eq:jmm}. This completes the proof of the Proposition \ref{lambdasteer}. 

It is important to note here that the existence of states associated with ontic variable $\lambda$ for the local statistics do not necessarily imply that an ontological model exists for the concerned statistics. Next, we  use  the Proposition \ref{lambdasteer} to establish the fact that indeed the existence of such an ontological model has an one-to-one correspondence with the existence of ontic state of the form of Eq. \eqref{eq:steer}. This is captured through the following theorem.

\begin{theorem}{}
 Bob's measurement statistics collected on measuring Alice's remotely prepared states at his wing satisfy generalised non-contextuality if and only if Alice's measurements $A_{x}$s are compatible. 
\end{theorem}

\textit{Proof:} We first prove the `only if' part, i.e., the ontological model of the operational theory corresponding to the local subsystem of Bob is generalised noncontextual \textit{only if} the measurements at Alice's wing are compatible. For `if' part, we show that  if Alice's observables are compatible then Bob's local statistics can be described by the generalised noncontextual ontological model.

To prove the `only if' part, let us consider that the joint statistics of Alice and Bob can be written in a GPT as,

 \begin{eqnarray}\label{steer2}
     p(a_{x},b_{y}|A_{x},B_{y}) &=& \langle \big(e_{\tiny A_{x}}^{\tiny a_{x}} \otimes e_{\tiny B_{y}}^{\tiny b_{y}}\big), \omega_{AB}\rangle
 \end{eqnarray}

Note that the no-signalling principle demands that the remote preparations at Bob's wing due to Alice's measurements must be operationally equivalent. Then, recalling the discussion of Sec. \ref{OntGPT}, the ontological model corresponding to the GPT describing the statistics of Bob's system must exist if,

\begin{eqnarray}\label{existence}
   &&\forall \ \bar{\xi}_{e^{b_{y}}_{B_{y}}} (\lambda), \ \ \   \bar{\mu}_{\omega_{A_{x}}}(\lambda)=\bar{\mu}_{\omega_{A'_{x}}}(\lambda)=\bar{\mu}(\lambda),  and  \nonumber \\
      &&\forall b_{y}, \bar{\mu}_{\omega_{A_{x}}}(\lambda),   \  B_{y} \equiv B'_{y}, \ \bar{\xi}_{e^{b_{y}}_{B_{y}}}(\lambda)=\bar{\xi}_{e^{b_{y}}_{B'_{y}}}(\lambda)  \ 
\end{eqnarray}
   Now, with the existence of such states and effects, the joint probability of Eq. \eqref{steer2} can be written as,

\begin{eqnarray}\label{steer3}
     p(a_{x},b_{y}|A_{x},B_{y}) &=& \sum_{\lambda} p(a_{x}|A_{x}) p(\lambda|a_{x},A_{x}) p(b_{y}|\lambda,B_{y}),
\end{eqnarray}

where the probabilities  $p(b_{y}|\lambda,B_{y}) = \bar{\xi}_{e^{b_{y}}_{B_{y}}}(\lambda) = \langle \omega^{\lambda}, e_{B_{y}}^{b_{y}} \rangle$, and $p(a_{x}|A_{x}) = \langle (e^{a_{x}}_{A_{x}}\otimes u_{B}), \omega_{AB}\rangle $. Further, from the
Bayes' rule one gets, $p(a_{x}|A_{x})p(\lambda|a_{x},A_{x})= p(\lambda|A_{x})p(a_{x}|\lambda,A_{x})$. Also, using Eq. \eqref{existence} we can write $p(\lambda|A_{x})= \bar{\mu}_{\omega_{A_{x}}}(\lambda) = \bar{\mu}(\lambda)$ such that Eq. \eqref{steer3} can be finally written as,

\begin{eqnarray}
    p(a_{x},b_{y}|A_{x},B_{y}) 
      &=& \sum_{\lambda} \bar{\mu}(\lambda) p(a_{x}|\lambda,A_{x})\langle \omega^{\lambda}, e_{B_{y}}^{b_{y}} \rangle
\end{eqnarray}

Clearly, in a no-signalling theory such joint statistics can be obtained when the local state at Bob's wing is of the form of Eq. \eqref{eq:steer}. Therefore, we established that the existence of an ontological model of the GPT describing Bob's local statistics implies that the states admits the decomposition in Eq. \eqref{eq:steer}. This in turn implies that Alice's measurements $A_x$s are compatible through the Proposition \ref{lambdasteer} . As already proved in  Theorem \ref{spekktheor} the existence of ontological model of GPT corresponding to an operational theory implies that the ontological model of that operational theory to be generalised noncontextual. Thus, the above argument along with Theorem \ref{spekktheor} leads the inference that  Bob's local statistics is generalised noncontextual only if Alice's measurements are compatible. Now, we  prove the converse implication.

The proof of "if" part first demands establishing the fact that nonexistence of the ontological model for the GPT describing Bob's local statistics implies that Bob's steered states cannot be written in the form given by Eq.\eqref{eq:steer}. Let us  assume that the ontological model for the GPT corresponding to Bob's local subsystem do not exist. This demands that either at least for some $ \bar{\xi}_{e^{b}_{M}} (\lambda)$, $\bar{\mu}_{\omega_{A_{x}}}(\lambda) \neq \bar{\mu}_{\omega_{A_{x'}}}(\lambda)$, and/or at least for some $ \bar{\mu}_{\omega_{A_{x}}}(\lambda) $, $\bar{\xi}_{e^{b}_{M}}(\lambda) \neq \bar{\xi}_{e^{b}_{M'}}(\lambda)$. In fact, nonexistence of any one of them will suffice. We can then write the joint statistics of Alice and Bob for two different measurements performed at Alice's wing as,

\begin{eqnarray}\label{nonequi}
    p(a_{x},b_{y}|A_{x},B_{y}) &=& \sum_{\lambda} p(\lambda|A_{x}) p(a_{x}|\lambda,A_{x})\langle \omega^{\lambda}, e_{B_{y}}^{b_{y}} \rangle \nonumber \\
 p(a_{x'},b_{y}|A_{x'},B_{y}) &=& \sum_{\lambda} p(\lambda|A_{x'}) p(a_{x'}|\lambda,A_{x'})\langle \omega^{\lambda}, e_{B_{y}}^{b_{y}} \rangle  
\end{eqnarray}

Now the assumption that for some $ \bar{\xi}_{e^{b}_{M}} (\lambda)$ for which $\bar{\mu}_{\omega_{A_{x}}}(\lambda) \neq \bar{\mu}_{\omega_{A_{x'}}}(\lambda)$ is equivalent to $p(\lambda|A_{x})\neq p(\lambda|A_{x'})$. This means, when ontological non-equivalence (in other word non-existence of ontological model itself) is assumed for operationally equivalent preparations, one cannot attribute independence of $\lambda$ for different preparation contexts. Therefore, Bob's local state admitting the decomposition of the form in Eq. \eqref{eq:steer} cannot produce the statistics given by Eqs. \eqref{nonequi}. Hence, existence of local states of the form Eq. \eqref{eq:steer} implies the existence of an ontological model for the GPT that reproduces Bob's local  statistics. Finally, using the Theorem \ref{spekktheor} and the  Proposition \ref{lambdasteer}, we can then ascertain that Bob's local state produces a generalised noncontextual statistics if the  Alice's measurements are compatible. This concludes the proof.

\begin{corollary}
\label{onetoone}
 A witness that is necessary for capturing  compatibility of Alice's  observables is also necessary for the existence of generalised noncontextual ontological model of remotely prepared states at Bob's wing.  
\end{corollary}

\begin{figure}
	\centering
	\includegraphics[scale = 0.4]{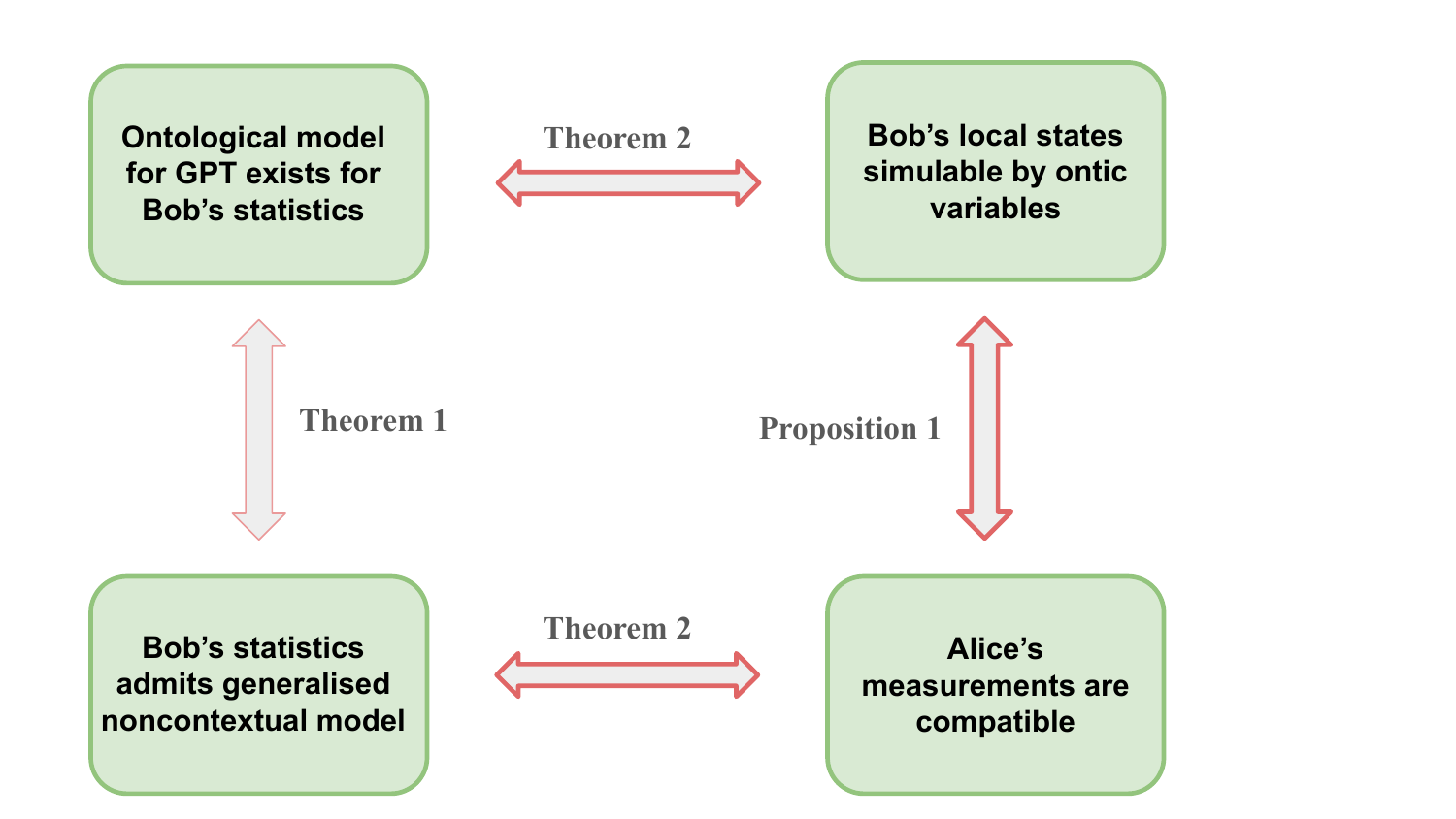}
	\caption{Pictorial depiction of main results. Dark red arrows are findings of the current paper while the blur arrow indicates the result of ref. \cite{spekkens21}.}
	\label{fig:seqd}
\end{figure}

We note that Corollary \ref{onetoone} is crucial as it sets the foundation of another key result which is discussed hereafter. The Corollary \ref{onetoone} also leads us to present a necessary condition for Alice's measurement compatibility (or generalised noncontextuality of Bob's local statistics) which is captured in the following theorem.
\begin{theorem}{}
 In any no-signalling GPT,  if Alice randomly performs measurement of $A_{x}$, with $x\in\{1,2,...,N\}$ on her subsystem and Bob randomly performs measurement of  $B_{y}$, with $y \in \{1,2,...,2^{N-1}\}$ on his subsystem then the necessary condition for $N$-wise compatibility of Alice's measurements is 

\begin{equation}\label{jgm}
    \mathcal{B}_{N}= \max_{B_{y}}\Big(\sum_{y=1}^{2^{N-1}}\lvert \sum_{x=1}^{N} (-1)^{l_{y}^{x}}\langle A_{x} \otimes B_{y} \rangle -\Delta(N)\rvert\Big) \leq 2^{N-1} 
\end{equation}
where $l_{y}^{x}$ is the $x^{th}$  bit sampled from $y^{th}$ string from the bit-string $l\in\{0,1\}^n$ having first entry $0$ and $\Delta(N)$ is a function of $N$-wise joint probabilities. 
\end{theorem}

\textit{Sketch of the proof:} As discussed earlier, if there exists a joint effect given by Eq. \eqref{eq:JM} whose suitable fine-grained version recovers all the effects in Eq. \eqref{eq:jmm} corresponding to Alice's measurements then such measurement are necessarily compatible. This implies that there exists joint probabilities of the form $p(a_{1},a_{2},...,a_{N}|A_{1},A_{2},....,A_{N})$ for the outcomes of such measurements. Further, due to the no-signalling condition, Alice's measurement statistics is independent of  Bob's  measurement choice $B_{y}$, i.e.,  

\begin{eqnarray}
\label{eq:nsig}
    &&p(a_{1},a_{2},...,a_{N}|A_{1},A_{2},....,A_{N}) \nonumber\\ &&=p(a_{1},a_{2},...,a_{N},b_{y}|A_{1},A_{2},....,A_{N},B_{y}) \nonumber\\
    &&+p(a_{1},a_{2},...,a_{N},-b_{y}|A_{1},A_{2},....,A_{N},B_{y})
\end{eqnarray}

Using those two assumptions (the joint measurability of Alice's observables and operational locality), we derive the  inequality in Eq. \eqref{jgm}. The explicit proof is given in Appendix A.

\begin{figure}
	\centering
	\includegraphics[scale = 0.45]{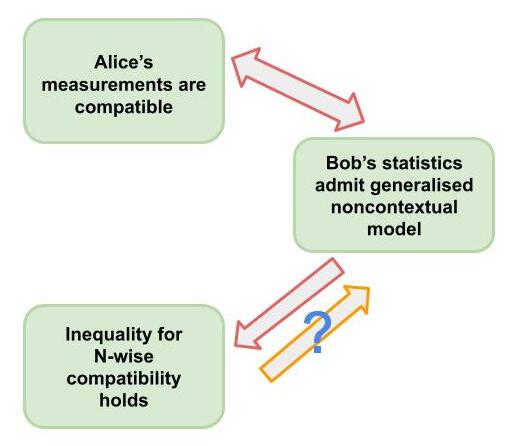}
	\caption{Schematic of the relation between status of $N$-wise compatibility of Alice's measurement and possibility of generalised noncontextual explanation of Bob's local subsystem.}
	\label{fig:seqd}
\end{figure}

We note again that the derivation of incompatibility witness in Eq. \eqref{jgm} is valid for any non-signalling GPT. The higher the violation of the inequality in a particular GPT, the higher incompatibility the concerned theory possesses. Thus, to determine the degree of incompatibility the quantum theory possesses, we need to derive the maximum quantum value of the functional $\mathcal{B}_{n}$. In the following, we derive the optimal quantum value of $\mathcal{B}_{n}$ without referring the dimension of the quantum system. Moreover, we provide the nature the observables required for achieving the maximum quantum bound.

\section{Optimal measurement incompatibility allowed in quantum theory}
 Consider that Alice's measurements are described by unbiased POVMs (smeared versions of projective measurement) of the form $A_x\equiv \Big\{A_{\pm|x} \ \big| \ E_{\pm|x}=\frac{1}{2}(\mathbb{I}\pm \eta A_x)\Big\}$, where $\eta \in [0,1]$ is the unsharpness parameter. Hence, effective observable of Alice is $\eta  A_{x}$. We note that for mutually-maximally incompatible quantum observables $\Delta(N)=0$. Thus, if we assume the same unsharpness parameter $\eta $ for each of the observable $A_{x}$, we can then write the quantum value of the functional $\mathcal{B}_{N}$ as,
 \begin{equation}
 \label{functional}
     (\mathcal{B}_{N})_{Q}=\eta \sum_{y=1}^{2^{N-1}} \rvert\sum_{x=1}^{N} (-1)^{l_{y}^{x}}\langle A_{x} \otimes B_{y} \rangle\rvert.
\end{equation}
To optimize $(\mathcal{B}_{N})_{Q}$ we use an elegant sum-of-square approach \cite{pan20} as explicitly described in the following. This implies showing  $(\mathcal{B}_{n})_{Q}\leq \beta_{N}$  for all possible quantum states $\rho_{AB}$ and measurement observables $A_{x}$ and $B_{y}$, where  $\beta_{N}$ is the upper bound. This is equivalent to showing that there is a positive semidefinite operator $\gamma_{N}\geq 0$, that can be expressed as $\langle \gamma_{N}\rangle_{Q}=\beta_{N}-(\mathcal{B}_{n})_{Q}$. We prove this by considering a set of suitable positive operators $L_{y}$ which is polynomial functions of   $A_{x}$ and $B_{y}$, so that,
\begin{align}
\label{gamma}
		\gamma_{N}=\sum\limits_{y=1}^{2^{N-1}}  \frac{\omega_{N,y}}{2}(L_y)^{\dagger}L_{y}
\end{align}
 where $\omega_{N,y}$ is positive semi-definite and to be specified shortly. It is then straightforward that the maximum value of $(\mathcal{B}_{n})_{Q}$ is obtained when $\langle \gamma_{N}\rangle_{Q}=0$, implying that 

\begin{align}
\label{mm}
	Tr[L_{y} \ \rho_{AB}]=0.
\end{align}

Now, without any loss of generality we can define $L_{y}$ as

\begin{align}
\label{mi}
	L_{y}=\frac{1}{\omega_{N,y}}\sum\limits_{x=1}^{N} (-1)^{y^x} A_{x} -B_{y}
\end{align}
where $	\omega_{N,y}=||\sum\limits_{x=1}^{N} (-1)^{y_x} A_{x}||$ is used for normalization.  Plugging Eq. (\ref{mi}) into Eq. (\ref{gamma}) and by noting that $A_{x}^{\dagger} A_{x}=B_{y}^{\dagger} B_{y}=\mathbb{I} $, we get

\begin{align}
\langle \gamma_{N}\rangle_{Q}=-(\mathcal{B}_{n})_{Q} + \sum\limits_{y=1}^{2^{N-1}}\omega_{N,y}
\end{align}
The maximum quantum value of $(\mathcal{B}_{N})_{Q}$ can be obtained when $\langle\gamma_{N}\rangle_{Q}= 0$, which in turn provides 

\begin{align}
\label{optbn}
(\mathcal{B}_{N})_{Q}^{max} &= \underset{A_{x}}{max}\left(\sum\limits_{y=1}^{2^{N-1}}\omega_{N,y}\right)\\
\nonumber
&=\underset{A_x}{max}\left(\sum\limits_{y=1}^{2^{N-1}}||\sum\limits_{x=1}^{N} (-1)^{y_x} A_{x}||\right)
\end{align}

To maximize $(\mathcal{B}_{N})_{Q}$, we use the concavity inequality, i.e., 
\begin{align}
\label{concav}
	\sum\limits_{y=1}^{2^{N-1}}\omega_{N,y}\leq \sqrt{2^{N-1} \sum\limits_{y=1}^{2^{N-1}} (\omega_{N,y})^{2}},
\end{align}
where the equality sign holds for Eq. (\ref{concav}) when $\omega_{N,y}$s are equal for each $y$.  The quantity $\omega_{N,y}$ can explicitly be written as 

\begin{align}
\label{omega}
	\omega_{N,y}&= \Big[ n+ \{(-1)^{y_1} A_{1}, \sum\limits_{x=2}^{N}(-1)^{y_x} A_{x}\} \\
	\nonumber
	&+\{(-1)^{y_2} A_{2}, \sum\limits_{x=3}^{N}(-1)^{y_x} A_{x}\} +........ \\
	\nonumber
	&+\{(-1)^{y_{N-1}} A_{N-1}, (-1)^{y_n} A_{N}\}\Big]^{-1/2}
\end{align}
where $\{,\}$ denotes the anti-commutation. It is then evident from Eq. \eqref{omega} that $\omega_{N,y}$s are equal to each other only when $A_x$s form a set of mutually anti-commuting operators. This in turn gives  $\omega_{N,y}=\sqrt{N}$ and consequently  from Eq. (\ref{optbn}) and \eqref{functional} the optimal quantum value of $(\mathcal{B}_{N})_{Q}$ is derived as,

\begin{align}
\label{opt}
(\mathcal{B}_{N})_{Q}^{max}= \eta 2^{N-1}\sqrt{N}
\end{align}
Since the cardinality of the set of mutually anti-commuting operators is always upper bounded for a fixed dimension one needs at least $d=2^{\lfloor n/2\rfloor}$ dimensional local Hilbert space to achieve the optimal quantum bound. Here we note that, in addition to the anti-commutation we already have another constraint for the local observables at Alice's side, namely, $A_{x}^{\dagger} A_{x}=\mathbb{I}$. Thus, more formally, we assert that the maximum violation of inequality Eq. \eqref{jgm} for any arbitrary $N$ is achieved when the following relation is satisfied among the measurements at Alice's side,

\begin{equation}
\label{anticom}
    \forall x, x', \ \ A_{x}A_{x'} +  A_{x'}A_{x}= 2\delta_{x,x'} \mathbb {I},
\end{equation}

This relation is satisfied when the operators $A_{x}$'s are the generators of a Clifford algebra \cite{clifford}. It is known that in any dimension $d$ a maximum of $d^{2}-1$ mutually anti-commuting dichotomic observables exist \cite{clifford}. Then for any value of $N$ there exists an upper bound on the dimension of the quantum system.

It is now straightforward to argue from Eq. \eqref{opt} that for a set of an arbitrary $N$  number of Alice's unsharp observables having the same unsharpness parameter $\eta$, the necessary condition for having $N$-wise compatibility is  $\eta\leq\frac{1}{\sqrt{N}}$. Then, according to Definition \ref{defeta}, quantum theory is simply a $N^{\frac{1}{\sqrt{N}}}$-incompatible theory. 

We note that the necessary condition for $N$-wise compatibility for unsharp POVMs derived above matches the necessary condition for $N$-wise compatibility of the same class of POVMs derived in \cite{kunjwal14,kunjwal14g}. However, while deriving the necessary condition they have used operator space structure of quantum theory itself. In contrast, staring from our theory-independent necessary condition derived in Eq.(\ref{jgm}), we reproduced their necessary condition by  inserting the operator formalism of quantum theory while optimizing the functional $(\mathcal{B}_{N})_{Q}$ in Eq. \eqref{functional}.  

Thus, our inequality can be cast as a witness for incompatibility whose violation quantifies the amount of incompatibility a set of observables possess. It is already shown that for a set of mutually anti-commuting observables the derived inequality can be violated maximally. Such a set forms the maximal incompatible set of observables in quantum theory, as already established in literature \cite{winter08}. However, in any other no-signalling theory there may not exists observables equivalent to something like a mutually anti-commuting set. For such theories other different values of unsharpness parameter $\eta^{\ast}$ (characteristics of that particular theory) is necessary to satisfy the constructed inequality for any arbitrary $N$, thereby possessing different degree of incompatibility as per the Definition \ref{defeta}.

%Moreover, supplementing this in \cite{kunjwal14} they have also shown (explicitly exemplified in Appendix B for $N=3$) that $\eta\leq\frac{1}{\sqrt{N}}$ is also a sufficient condition for the $N$-wise compatibility.

\section{Geometric implications}

So far, we have established the one-to-one connection between generalised noncontextuality of the steered subsystem in one wing of a nontrivially correlated bipartite system and the compatibility of measurements performed on the subsystem at other wing. Moreover, we have also formulated the necessary condition for the $N$-wise compatibility in the context of such bipartite scenario. At this point, it is essential to analyse what geometrical restrictions this $N$-wise compatibility provides to the GPT describing the steered states produced due to such measurements. In order to do this we begin with discussing some geometrical properties of a GPT as follows:

\begin{definition}{Simplicial GPTs:} \label{simplicialdef}
A GPT of dimension $d$ is called simplicial if its state space $\Omega$ can be considered as a simplex $\mathcal{S}$ consisting of exactly $d$ vertices and intrinsic dimension $d-1$ living inside a real vector space $\mathbb{R}^{d}$, and also the effect space $\Xi$ constitutes hypercube $\mathcal{S}^*$ (dual to the simplex) in dimension $d$ and having $2^{d}$ vertices.
\end{definition}

 According to Definition \ref{simplicialdef} the simplex $\mathcal{S}$ lies in the real vector space $\mathbb{R}^{d}$ associated with an Euclidean topology. Accordingly for the set of points $\hat{X}\in\{\hat{s}_{1},\hat{s}_{2}...\hat{s}_{N}\}$ as its extreme points there exist affine functions $\hat{s}_{j}' : \mathcal{S}\rightarrow \mathbb{R}$, with $j\in \{1,2...N\}$, such that $\langle \hat{s}_{i}, {s}_{j}'\rangle=\delta_{i,j}$. Any point in the convex hull (represented as $\textbf{Con}(\{s_{i}\})$) of the points $\hat{X}\in\{\hat{s}_{1},\hat{s}_{2}...\hat{s}_{N}\}$ can be represented as, $\vec{r}=\sum_{i} q_{i} \hat{s}_{i}$, with $0 \leq q_{i} \leq 1 $. Thus, the function $\hat{s}_{j}'$ must be positive and all $r\in \textbf{Con}(\{s_{i}\})$ are valid states representing the simplicial GPT, $\mathcal{S}:=\textbf{Con}(\{s_{i}\})$. On the other hand, the effect space must be a hypercube $\mathcal{S}^*$ of intrinsic dimension same as the dimension of the state space consisting of $2^{d}$ vertices representing the extremal effects. Thus, the effect space must lie inside the convex hull of points $\{\sum_{i \in \mathcal{T} } \hat{s}_{i} \}$, with $\mathcal{T} \subseteq X$.
 
 In literature \cite{barrett07} the simplicial GPTs are referred to as the classical due to the following reason. If the extremal states are the representative of comprehensively distinct states of the system then any legitimate pure state should correspond to those extremal states with an extra proviso that any legitimate mixed state must portray the epistemic uncertainty about the extremal states. Thus every mixed state should be uniquely determined by the convex decomposition of the pure states. In discrete geometry, simplex is the only regular geometrical structure that has this specific property that any point inside the simplex can be uniquely decomposed into the points representing its vertices. However, there is no reason to think that every classical GPT would be simplicial. Nevertheless there are GPTs that are not simplicial, but they can embadded into a simplicial GPT. Such GPTs are defined as as follows.

\begin{definition}{Simplex-embeddable GPTs:} \label{def1}
A particular GPT $\zeta = (\Omega,\Xi)$ over space $\mathcal{V}$ is simplex-embeddable if there exists a simplicial GPT $\bar{\zeta} = (\bar{\Omega},\bar{\Xi})$ over space $\mathcal{\bar{V}}$ of some dimension $d$, and linear maps $\mathcal{M}_{\Omega}$, $\mathcal{M}_{\Xi} :$ $\mathcal{V}\rightarrow \mathcal{\bar{V}}$ such that, 
\begin{eqnarray}
    \mathcal{M}_{\Omega}(\Omega) &\subseteq& \bar{\Omega}, \nonumber \\
    \mathcal{M}_{\Xi}(\Omega) &\subseteq& \bar{\Xi}, \nonumber \\
   and \  \langle e , \omega \rangle &= & \langle \mathcal{M}_{\Xi}(e) , \mathcal{M}_{\Omega}(\omega) \rangle, \ \ \ \ \forall \ \omega \in \Omega, e \in \Xi
\end{eqnarray}
\end{definition}

Thus there may exist GPTs that are simplex-embeddable but not simplicial. Such GPTs can then be treated as classically simulable rather than classical as noted in \cite{shahandeh21}. Moreover, it is argued in \cite{spekkens21} that a simplex-embeddable GPT that respects \textit{no-restriction hypothesis} \cite{janotta13} must be a simplicial GPT. In this paper however, we do not restrict ourselves only to the GPTs that respects the \textit{no-restriction hypothesis} and consider the notion of simplicticity and simplex-embeddability to be two distinct layer of classicality.

%\begin{widetext}

\begin{figure}
	\centering
	\includegraphics[scale = 0.15]{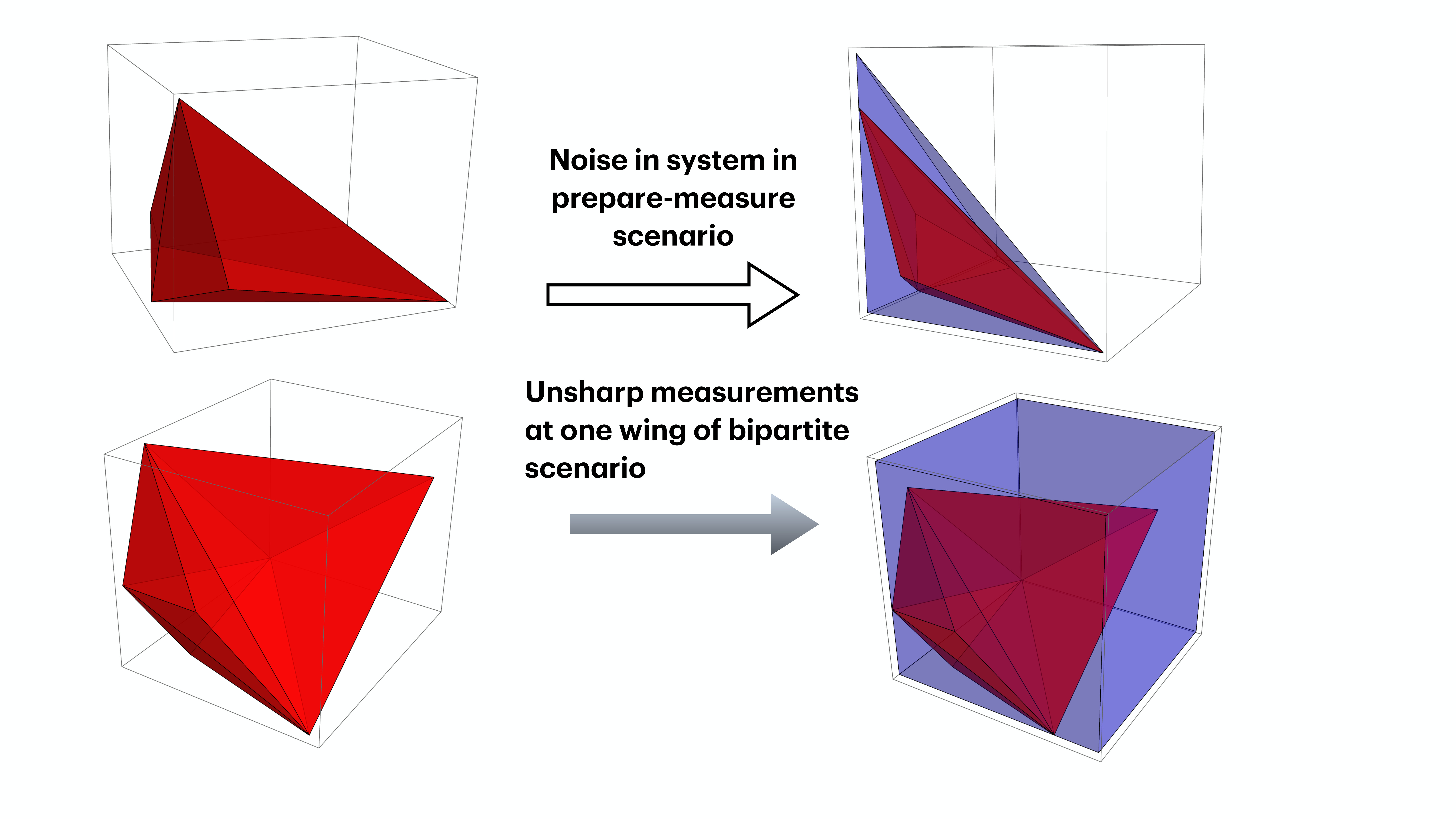}
	\caption{Schematic diagram of state and effect spaces represented in three dimensional Euclidean space. While the left side of the arrow depicts an arbitrary shaped state and effects, the right side shows that the state and effect space transformation in order to embedding them into a simplex and hypercube respectively. The no-fill arrows show the result obtained in \cite{rossi23,selby24} that suitable noise in state viable of simplex-embeddability, while the gray arrow shows the result obtained in this paper namely the unsharpness in measurement is another viable parameter of simplex-embeddability.}
	\label{fig:seqd}
\end{figure}

%\end{widetext}
%\begin{figure}
	%\centering
	%\includegraphics[scale = 0.4]{sim.jpg}
	%\caption{Schematic of how a non-simplicial GPT can be embedded into a simplicial GPT by allowing noise in state and effects respectively.}
	%\label{fig:seqd}
%\end{figure}

Now, we discuss the geometrical properties of the GPT describing Bob's local statistics obtained from the remotely prepared systems discussed in the preceding sections in terms of its connections with the Alice's compatible measurements. For this, we recall the result of ref. \cite{spekkens21} that the ontological model of an operational theory is generalised noncontextual \emph{iff} the GPT describing the operational statistics is simplex-embeddable. It is already clear from previous section that in a bipartite scenario with a nontrivially correlated shared state, if Alice's measurements are all compatible then Bob's statistics can be reproduced by a generalised noncontextual model. By connecting the result of ref. \cite{spekkens21} and the results obtained in this paper we prove the following theorem. 

\begin{theorem}{}
In a bipartite scenario with a nontrivially correlated shared state $\Omega_{AB}$, local statistics of Bob's steered subsystem can be explainable by a simplex-embeddable GPT if and only if Alice's measurements $\{A_{x}\}_{x=1}^{N}$ are  compatible. 
\end{theorem}

\textit{Proof:} Consider the state space of a GPT is a unit $d$-simplex $\mathcal{S}$ with the effect space a unit hypercube $\mathcal{S}^{*}$ dual to the simplex lying in $\mathbb{R}^{d}$. This then forms a simplicial GPT according to Definition \ref{simplicialdef}. Now, the proof is very straightforward from ref. \cite{spekkens21} when one identifies the $\lambda \in \Lambda$ as the ontic states describing the steered preparations. Due to the positivity of points $\{s_{i}\}$ one easily identify the vertices of the simplex $\mathcal{S}$ is isomorphic with each of the variable $\lambda \in \Lambda$ and so with the elements of the set $a=\{a_{x}\}_{x=1}^{N}$. For instance, we consider the set $a$ as points on a vector space $\mathbb{R}^{\Lambda}$ forming $n(a)$ number of vertices of the simplex $\mathcal{S}$, with $n(a)$ being the cardinality of the set $a$. 

Next, the real vector space $\mathbb{R}^{\Lambda}$ has a one-to-one correspondence with the GPT vector space $\mathcal{V}$ such that any vector $\vec{v} \in \mathcal{V}$ is associated with a function $\vec{g}_{v}\in\mathbb{R}^{\Lambda}$ with $\vec{g}_{v}=\sum_{i}g_{v}^{i}\hat{s}_{i}$. Therefore, the existence of the set $a$ signifies that the GPT state space of the steered preparations can be mapped to the state state space of a simplicial GPT.

Along the same line, we argue that the vertices of $\mathcal{S}^{*}$, namely the points $\sum_{i \in \mathcal{T} } \hat{s}_i $, with with $\mathcal{T} \subseteq X$ can be considered as the response functions of the ontological model of the GPT compatible with Bob's local statistics when Alice's measurements are compatible. Strictly speaking, there exists a set of functions $\hat{s}_{j}': \mathcal{S}^{*} \rightarrow \mathbb{R}$ such that, these functions uniquely identifies any of the extremal points of  $\mathcal{S}^{*}$  as, $\sum_{i\in\mathcal{T}}\langle \hat{s}_{i}, \hat{s}_{j}'\rangle= c_{i}$. These functions directly corresponds to the response functions used in the definition of the ontological model of a GPT. Also, the inner product between any two vectors $\vec{v},\vec{v}'\in \mathcal{V}$ can be written as the dot products of functions $\vec{g}_{v},\vec{g}_{v'}\in \mathbb{R}^{\Lambda}$ as,

\begin{equation}
    \langle \vec{v},\vec{v}' \rangle= \vec{g}_{v}\cdot \vec{g}_{v'} = \sum_{i,j}g_{v}^{i}g_{v'}^{j}\delta_{ij}
\end{equation}

It is then evident that when Alice's measurements are compatible, one can always assign an ontological model of the GPT compatible with Bob's local statistics,  thereby assures that such GPT can be embedded into a simplicial GPT. Equivalently, each simplex embeddable GPT description of Bob's local statistics assures that the measurements used to remotely prepare its states must be compatible.

\section{Summary and Conclusions}

In sum, we demonstrated a hitherto unexplored connection between measurement incompatibility and generalised noncontextuality in any arbitrary no-signalling GPT. For this, we considered a bipartite Bell scenario involving a non-trivially correlated system shared between two specially separated parties Alice and Bob. We proved that Bob's local statistics can be reproduced by a generalised noncontextual model  if an only if Alice's measurements $\{A_{x}\}_{x=1}^{N}$  are compatible,where $N$ is arbitrary.  

We note here that a number of studies have been made toward establishing a similar connection between incompatibility and various forms of quantum correlations, viz. nonlocality \cite{bene18,hirsch18} steering \cite{uola15,uola14}, preparation and measurement noncontextuality \cite{tavakoli20,singh23}. In this work, we have moved a step forward by establishing the connection between measurement incompatibility and generalised noncontextuality in any no-signalling operational theory, and thereby quantifying the maximum allowed incompatibility of a theory determined by the generalised contextuality.

Further, we formulated a non-trivial inequality solely by considering the $N$-wise compatibility of Alice's measurements $\{A_{x}\}_{x=1}^{N}$ and the operational locality in the GPT. This inequality forms a necessary condition for Alice's compatibility and generalised noncontextual description for Bob's statistics. Thus, the violation of that inequality implies $N$-wise incompatibility of the Alice's measurements and establishes generalised contextuality of Bob's statistics. Note however that the maximal violation of the inequality varies depending on which specific no-signalling GPT one considers. We introduced a notion of $N^{\eta^{*}}$-compatible GPT which quantifies the degree of incompatibility of the GPT concerned. 

For quantum theory, by using an elegant optimization technique we derived the optimal quantum violation of the inequality, which in turn gives the necessary condition for $N$-wise compatibility when $\eta \leq 1/\sqrt{N}$. Thus, the maximum violation of the inequality is limited by the maximum generalised contextuality that the theory allows. Following Definition \ref{defeta}, we argue that any theory for which the necessary condition of $N$-wise compatibility is $\eta<1/\sqrt{N}$ (or $\eta > 1/\sqrt{N}$) is more (or less) incompatible theories than quantum theory, and accordingly entails more (or less) generalised contextuality. Thus, if one assumes that quantum theory is the governing theory of nature then the aforementioned two theories do not exist.

Furthermore, we have discussed the geometrical implications of  the $N$-wise compatibility in a GPT. It has recently been proved in \cite{spekkens21} that a GPT is simplex-embeddable if and only if the ontological model corresponding to the operational theory this GPT concerns is generalised noncontextual. We demonstrated that any set of Alice's observables that is used to remotely prepare Bob's states is compatible if and only if Bob's local statistics can be described in terms of a simplex-embeddable GPT. In \cite{rossi23,selby24}, it is shown that any preparation whose corresponding statistics cannot be described in terms of a simplex-embeddable GPT can be made simplex-embeddable by allowing dephasing noise to the states. In contrast to Refs.\cite{rossi23,selby24}, where prepare-measure scenario was considered, we argued  that in our bipartite Bell scenario,  Alice's unsharp measurements remotely prepares noisy states so that a simplex-embeddable GPT explanation exists (which is otherwise not simplex-embeddable). Obviously, the degree of unsharpness parameter  needs to be bounded. In quantum theory, simplex-embeddability exists for $\eta \leq 1/\sqrt{N}$.

In recent work \cite{selby23c,selby23}, it is argued that measurement incompatibility is not a prerequisite to reveal generalised contextuality in the preparation and measurement scenario. Although it apparently seems, there is no contradiction with the results of our paper. We claimed that the compatibility of Alice's measurements used to steer Bob's states is necessary and sufficient to reveal the generalised contextuality of the steered preparations. On the other hand, the claim of \cite{selby23c,selby23} refers only to the connection between the measurements used to obtain the statistics from a single system and the possibility of revealing the generalised contextuality. Therefore, the domain of applicability of the results obtained in \cite{selby23c,selby23} and this paper is different.

Finally, we may consider a more general definition of the incompatibility of a theory in terms of the compatibility structure that exists in such a theory. In fact, it would be interesting to find connections between non-classical correlations exhibited by an arbitrary operational theory and other layers of non-classicality \cite{layers16,layers21} exist for measurements apart from the $N$-wise incompatibility. This would enable one to explore the uniqueness of quantum theory compared to other no-signaling theories. We note that, while discarding other class of incompatible theories in terms of our definition, we have assumed the superiority of generalized contextuality of quantum theory. It would be more intriguing if one can find a physical principle that can discard such theories and super-select the quantum theory. Such an attempt \cite{gonda18} was made which is particularly designed for the global joint measurability of sharp measurements through the Speaker's principle. Studies along those direction could be an exciting avenue for future research and thus calls for further study.\\

\emph{Acknowledgements:-}
S.M. acknowledges David Schmid for his comment on an earlier version of the manuscript. A.K.P acknowledges the support from the research grant SERB/MTR/2021/000908, Government of India. This work was also partly supported by Institute for Information and Communications Technology Planning and Evaluation (IITP) grant funded by the Korea government (MSIT) (RS-$2023-00222863$).

\appendix

\begin{section}{Derivation of necessary condition for $N$-wise compatibility}

 In order to derive the necessary condition for $N$-wise compatibility of $N$ dichotomic observables we consider a bipartite scenario featuring two specially separated observers Alice and Bob. The joint system is described by a state $\omega_{AB}$ in a GPT. Alice  randomly performs the measurement of dichotomic observables $A_{x}= \{e_{\tiny A_{x}}^{\tiny a_{x}}\}$, with $x \in \{1,...,N\}$, and Bob randomly performs the measurement of $B_{y}=\{e_{\tiny B_{y}}^{\tiny b_{y}}\}$, with $y \in \{1,...,2^{N-1}\}$. Here, $a_{x},b_{y}\in \{\pm 1\}$, denotes the outcomes of Alice and Bob respectively. The joint probability can then be written as,

 \begin{equation}
     p(a_{x},b_{y}|A_{x},B_{y})= \langle \big(e_{\tiny A_{x}}^{\tiny a_{x}} \otimes e_{\tiny B_{y}}^{\tiny b_{y}}\big), \omega_{AB}\rangle
 \end{equation}
and the correlation functions take the form,

 \begin{equation}
 \label{eq:corr}
     \langle A_{x} \otimes B_{y} \rangle = p(a_{x}=b_{y}|A_{x},B_{y})-p(a_{x}=-b_{y}|A_{x},B_{y})
 \end{equation}

with $\otimes$ is suitable tensor product specific to the mathematical structure of the particular GPT concerned.

 \begin{widetext}
The $N$-wise compatibility given by Eq.\eqref{nwise} implies the existence of the $N$-wise joint probability $p(a_{1},a_{2},....,a_{N}|A_{1},A_{2},....,A_{N})$. Then, from the no-signalling condition, the following relation holds.
\begin{eqnarray}
\label{eq:elegant1}
    p(a_{1}=a_{2}=....=a_{N}|A_{1},A_{2},....,A_{N}) &=& p(a_{1}=a_{2}=....=a_{N}=b_{1}|A_{1},A_{2},....,A_{N},B_{1}) \nonumber\\
    &&+p(a_{1}=a_{2}=....=a_{N}=-b_{1}|A_{1},A_{2},....,A_{N},B_{1})
\end{eqnarray}

 We also note that
 \begin{eqnarray}
 \label{eq:elegant2}
&&p(a_{1}=a_{2}=....=a_{N}=b_{1}|A_{1},A_{2},....,A_{N},B_{1})+p(a_{1}=a_{2}=....=a_{N}=-b_{1}|A_{1},A_{2},....,A_{N},B_{1}) \nonumber \\
&&\geq \lvert p(a_{1}=a_{2}=....=a_{N}=b_{1}|A_{1},A_{2},....,A_{N},B_{1})-p(a_{1}=a_{2}=....=a_{N}=-b_{1}|A_{1},A_{2},....,A_{N},B_{1}) \rvert
\end{eqnarray}

 Now, following Eq. \eqref{eq:corr}, a straightforward calculation gives,

 \begin{eqnarray}
 \label{eq:elegent3}
     &&p(a_{1}=a_{2}=....=a_{N}=b_{1}|A_{1},A_{2},....,A_{N},B_{1})-p(a_{1}=a_{2}=....=a_{N}=-b_{1}|A_{1},A_{2},....,A_{N},B_{1}) \nonumber \\
     && \ \ \ \ \ \ \ \ \ \ \ =\frac{1}{2^{N-1}}\langle \ (A_{1}+A_{2}+.....+A_{N})\ \otimes B_{1} \ \rangle -\Delta(N),
 \end{eqnarray}

where $\Delta(N)$ is a function of $N$-wise probabilities. Further, by using Eq. \eqref{eq:elegant1}-\eqref{eq:elegent3}, we can write,

\begin{equation}
\label{mod1}
    p(a_{1}=a_{2}=.....=a_{N}|A_{1},A_{2},....,A_{N}) \geq \frac{1}{2^{N-1}}\lvert \langle \ (A_{1}+A_{2}+.....+A_{N})\ \otimes B_{1} \ \rangle -\Delta(N)\rvert
\end{equation}

Following the similar steps, we derive $2^{N-1}-1$ more inequalities of the following forms.

\begin{eqnarray}
\label{mod2}
    p(a_{1}=-a_{2}=.....=a_{N}|A_{1},A_{2},....,A_{N}) & \geq & \frac{1}{2^{N-1}} \lvert \langle \  (A_{1}-A_{2}+A_{3}.....+A_{N})\ \otimes B_{2} \ \rangle -\Delta(N)\rvert\nonumber \\
     p(a_{1}=a_{2}=-a_{3}=.....=a_{N}|A_{1},A_{2},A_{3}....,A_{N}) & \geq & \frac{1}{2^{N-1}} \lvert \langle \ (A_{1}+A_{2}-A_{3}.....+A_{N})\ \otimes  B_{3} \ \rangle -\Delta(N)\rvert \nonumber \\ 
       ............... && \ \ \ ........ \ \ \ \ \ \ \ \ \ \ \ \ \ \ \ \ \ ........ \nonumber \\
     ............... && \ \ \ ........ \ \ \ \ \ \ \ \ \ \ \ \ \ \ \ \ \ ........ \nonumber \\
      ....... ........ && \ \ \ ........ \ \ \ \ \ \ \ \ \ \ \ \ \ \ \ \ \ ........ \nonumber \\
     p(a_{1}=-a_{2}=.....=-a_{N-1}=-a_{N}|A_{1},A_{2},....,A_{N}) & \geq & \frac{1}{2^{N-1}} \lvert \langle \ (A_{1}-A_{2}-A_{3}.....-A_{N-1}-A_{N})\  \otimes B_{2^{N-1}} \ \rangle-\Delta(N) \rvert 
\end{eqnarray}

The set of $2^{N-1}$ inequalities can  be written in a generalised form as,

\begin{equation}
\label{eq:gnrl}
    p\Big( a_{1}=(-1)^{l_{y}^{2}}a_{2}=.....=(-1)^{l_{y}^{N-1}}a_{N-1}=(-1)^{l_{y}^{N}}a_{N} \ | \ A_{1},A_{2},....,A_{N}\Big) \geq \frac{1}{2^{N-1}} \lvert \sum_{x=1}^{N}(-1)^{l_{y}^{x}} \langle A_{x} \otimes B_{y}\rangle -\Delta(N)\rvert.
\end{equation}

where $l^{x}_{y}$ is the $x^{th}$  bit sampled from the $y^{th}$ string from the bit string $l\in\{0,1\}^n$ that has the first entry $0$. Note that, for valid probabilities we must have, 

\begin{eqnarray}
\label{eq:nrml}
   \sum_{y=1}^{2^{N-1}}p(a_{0}=(-1)^{l_{y}^{1}}a_{1}=.....=(-1)^{l_{y}^{N-1}}a_{N-1}=(-1)^{l_{y}^{N}}a_{N} \ | \ A_{0},A_{1},....,A_{N})=1
\end{eqnarray}

\end{widetext}

Then by using Eq.\eqref{eq:gnrl} and Eq. \eqref{eq:nrml} we can then write the following inequality ,

   \begin{equation}
     \sum_{y=1}^{2^{N-1}}\lvert \sum_{x=1}^{N} (-1)^{l_{y}^{x}}\langle A_{x} \otimes B_{y} \rangle -\Delta(N)\rvert \leq 2^{N-1} 
\end{equation}
%By noting identity $|m+n| \leq |m|+|n|$, with $m,n$ being two arbitrary functions, we can write the above equation as,

 %\begin{equation}
   %\lvert \sum_{x=1}^{N}  \sum_{y=1}^{2^{N-1}} (-1)^{l_{y}^{x}}\langle A_{x} \otimes B_{y} \rangle -\Delta(N) \rvert \leq 2^{N-1} 
%\end{equation}

Finally, we get the necessary condition for the $N$-wise compatibility for Alice's observables as,

\begin{equation}
     \max_{B_{y}}\Big(\sum_{y=1}^{2^{N-1}}\lvert \sum_{x=1}^{N} (-1)^{l_{y}^{x}}\langle A_{x} \otimes B_{y} \rangle -\Delta(N)\rvert\Big) \leq 2^{N-1} 
\end{equation}
which is Eq. \eqref{eq:nsig} in the main text.

\end{section}
\section{Existing necessary condition for joint measurability in quantum theory implied by our inequality }

In order to discuss the necessary condition for the $N$-wise  compatibility of quantum measurements, let us take binary outcome observable $A_{x}$ where $A_{x} = \{E_{x}^{a_{x}}|E_{x}^{a_{x}} > 0; \sum_{a_{x}}E_{x}^{a_{x}}= \mathbb{I}\}$. As discussed, $A_{x}$s are jointly measurable if there exists global POVMs $E_{1,2....N}^{a_{1},a_{2}...a_{N}}$ such that,

\begin{equation}
	E_{x}^{a_{x}}=\sum_{\vec{m}/a_{x}}E_{1,2....N}^{a_{1},a_{2}...a_{N}}
\end{equation}

It is already established \cite{kunjwal14,kunjwal14g,kunjwal20} that for $N$ binary POVMs of the form, $E_{x}^{a_{x}}= \frac{\mathbb{I}+\eta a_{x}\Gamma_{x}}{2}$, where $\Gamma_{x}$ are the generators of Clifford algebra satisfying Eq. (\ref{anticom}), the necessary condition for $N$-wise compatibility is,

\begin{equation}
\label{nec1}
    \eta \leq \frac{1}{N}||\vec{r}\cdot \vec{\Gamma} ||
\end{equation}

where $\vec{r}= \{r_{1},r_{2},.....,r_{N}\}$, with $r_{x} \in a_{x}:=\{+1,-1\}$. Note that the value of $||\vec{r}\cdot \vec{\Gamma} ||=\sqrt{N}$, which in turn gives $\eta\leq 1/\sqrt{N}$.

On the other hand, we derived the necessary condition for compatibility of measurements of the above form ($\Delta(3)=0$) as,

\begin{equation}
\label{nec2}
    \eta \leq \frac{2^{N-1}}{   \max_{B_{y}}\Big(\sum_{y=1}^{2^{N-1}}\lvert \sum_{x=1}^{N} (-1)^{l_{y}^{x}}\langle A_{x} \otimes B_{y} \rangle \rvert\Big)\Big) }  
\end{equation}

Now, if we assume quantum theory as the operational theory, it is straightforward to see that by substituting   $A_{x}=\Gamma_{x}$,  the quantity $\max_{B_{y}}\Big(\lvert\sum_{x=1}^{N}  \sum_{y=1}^{2^{N-1}} (-1)^{l_{y}^{x}}\langle A_{x} \otimes B_{y} \rangle \rvert\Big) =2^{N-1}\sqrt{N} $ which is achieved when $A_{x}$ satisfies Eq. (\ref{anticom}). This in turn gives the same necessary condition $\eta\leq 1/\sqrt{N}$.

%In order to derive the necessary and sufficient condition for the joint measurability of $N$ arbitrary number of quantum measurements let us take binary outcome observable $A_{x}$ where $A_{x} = \{E_{x}^{m_{x}}|E_{x}^{m_{x}} > 0; \sum_{m_{x}}E_{x}^{m_{x}}= \mathbb{I}\}$. Now, all the $A_{x}$ are jointly measurable if there exists a global POVM $E_{1,2....N}^{m_{1},m_{2}...m_{N}}$ such that,

%\begin{equation}
	%E_{x}^{m_{x}}=\sum_{\vec{m}/m_{x}}E_{1,2....N}^{m_{1},m_{2}...m_{N}}
%\end{equation}

For a specific example, we show the above equivalence for $N=3$, by using qubit measurements. For $N$ binary qubit POVMs of the form, $E_{x}^{a_{x}}= \frac{\mathbb{I}+\eta a_{x}\hat{k}_{x} \cdot \hat{\sigma}}{2}$ the necessary condition Eq.\eqref{nec1} for N-wise Joint measurability can be equivalently written as,

\begin{equation}\label{necessaryk}
    \eta \leq \frac{1}{N} \max_{\vec{m}} \big|\big| \sum_{x=1}^{N} a_{x} \hat{k}_{x}\big|\big|
\end{equation}

with $\vec{a}= \{a_{x}\}_{x=1}^{N}$ and $a_{x}\in \{\pm 1\}$. Now, three observables, $A_{1}$, $A_{2}$ and $A_{3}$ are jointly measurable if,

\begin{eqnarray}
		E_{1}^{+}&&=E_{1,2,3}^{+++}+E_{1,2,3}^{++-}+E_{1,2,3}^{+-+}+E_{1,2,3}^{+--}  \nonumber\\ 
			E_{1}^{-}&&=E_{1,2,3}^{-++}+E_{1,2,3}^{-+-}+E_{1,2,3}^{--+}+E_{1,2,3}^{---}
\end{eqnarray} 
and similarly $E_{2}^{+}, E_{3}^{+}, E_{2}^{-}$ and $ E_{3}^{-}$ can be given equivalent representations.
From Eq. \eqref{necessaryk} the necessary condition for joint measurability of three anti-commuting observables becomes

\begin{equation} \label{eq:necessaryk}
	\eta \leq \frac{1}{3} \max_{x_{1},x_{2},x_{3}} \left( \big|\big|a_{x_{1}x_{2}x_{3}}\big|\big|\right) 
\end{equation}
where $a_{x_{1}x_{2}x_{3}}=\sum_{x=1}^{3}a_{x}\hat{k}_{x}$. Now, due to symmetry $\forall a_{1},a_{2},a_{3}$ we must have, $||a_{a_{1}a_{2}a_{3}}||=\sqrt{3}$. This in turns gives $ \max_{a_{1},a_{2},a_{3}} \{||a_{a_{1}a_{2}a_{3}}||\}=\sqrt{3}$. Then from Eq. \eqref{eq:necessaryk} we finally get that the joint measurability condition for three unbiased POVMs as $\eta \leq 1/\sqrt{3}$.

Now, let us re-write our derived necessary condition for compatibility for $N=3$ mutually-maximally incompatible observables ($\Delta(3)=0$) as,
\begin{eqnarray}\label{jm3}
    &&\mathcal{B}_{3}=\lvert \langle (A_{0}+A_{1}+A_{2})B_{0} \rangle \rvert + \lvert \langle (A_{0}-A_{1}+A_{2})B_{1} \rangle \rvert \nonumber \\
   && + \lvert \langle (A_{0}+A_{1}-A_{2})B_{2} \rangle \rvert+ \lvert \langle (A_{0}-A_{1}-A_{2})B_{3} \rangle \rvert \leq 4
\end{eqnarray}
In quantum theory, $(\mathcal{B}_{3})_Q^{max} = 4\sqrt{3}$ which can be obtained when $A_{0}=\sigma_{x}$, $A_{1}=\sigma_{y}$, and $A_{2}=\sigma_{z}$ with appropriate choice of $B_{y}$s. However, if Alice implements unsharp version of above measurements, from Eq. \eqref{jm3} we get, $\eta(\mathcal{B}_{3})_Q \leq 4$. Thus, the necessary condition  for compatibility comes out to be $\eta \leq 1/\sqrt{3}$.

\end{document}